\documentclass[sigconf,nonacm]{acmart}

\usepackage{smartref}
\usepackage{booktabs}
\usepackage{cleveref}
\usepackage{multirow}
\usepackage{listings}
\usepackage{algorithm}
\usepackage{algpseudocode}
\makeatletter
\newcommand\HL{%
	\gdef\lst@alloverstyle##1{%
		\fboxrule=0pt
		\fboxsep=0pt
		\colorbox{lightgray}{\strut##1}%
	}%
}

\newcommand\HLoff{%
	\xdef\lst@alloverstyle##1{##1}%
}

\newcommand{\code}[1]{\texttt{#1}}

\newcommand{\name}{TypeWriter}
\newcommand{\fb}{Facebook} % for double-blind review
\newcommand{\tool}[1]{\textsf{#1}}
\newcommand{\mypy}{\tool{mypy}}

\newcommand{\topk}{top-\emph{k}}

\renewcommand\footnotetextcopyrightpermission[1]{} % removes footnote with conference information in first column

\begin{document}

\title{\name: Neural Type Prediction with \mbox{Search-based Validation}}

\author{Michael Pradel}
\authornote{Work performed while on sabbatical at Facebook, Menlo Park.}
\affiliation{%
  \institution{University of Stuttgart}
}
\email{michael@binaervarianz.de}

\author{Georgios Gousios}
\authornotemark[1]
\affiliation{%
  \institution{Delft University of Technology}
}
\email{g.gousios@tudelft.nl}

\author{Jason Liu}
\email{jasonliu@fb.com}
\affiliation{%
 \institution{Facebook}
}

\author{Satish Chandra}
\email{schandra@acm.org}
\affiliation{%
 \institution{Facebook}
}

\begin{abstract}
Maintaining large code bases written in dynamically typed languages, such as JavaScript or Python, can be challenging due to the absence of type annotations: simple data compatibility errors proliferate, IDE support is limited, and APIs are hard to comprehend.
Recent work attempts to address those issues through either static type inference or probabilistic type prediction.
Unfortunately, static type inference for dynamic languages is inherently limited, while probabilistic approaches suffer from imprecision.
This paper presents \name{}, the first combination of probabilistic type prediction with search-based refinement of predicted types.
\name{}'s predictor learns to infer the return and argument types for functions from partially annotated code bases by combining the natural language properties of code with programming language-level information.
To validate predicted types, \name{} invokes a gradual type checker with different combinations of the predicted types, while navigating the space of possible type combinations in a feedback-directed manner.

We implement the \name{} approach for Python and evaluate it on two code corpora: a multi-million line code base at \fb{}
%that powers applications used by billions of people
 and a collection of 1,137 popular open-source projects.
% We show that \name{}'s type predictor achieves a precision of 64\% (91\%) and a recall of 52\% (68\%) in the top-1 (top-5) predictions, which clearly outperforms prior type prediction models.
We show that \name{}'s type predictor achieves an F1 score of 0.64 (0.79) in the top-1 (top-5) predictions for return types, and 0.57 (0.80) for argument types, which clearly outperforms prior type prediction models.
By combining predictions with search-based validation, \name{} can fully annotate between 14\% to 44\% of the files in a randomly selected corpus, while ensuring type correctness.
A comparison with a static type inference tool shows that \name{} adds many more non-trivial types.
\name{} currently suggests types to developers at \fb{} and several thousands of types have already been accepted with minimal changes.
%Overall, \name{} provides developers with an effective way to help with the transition to fully type-annotated code.
\end{abstract}

\maketitle

\section{Introduction}

Dynamically typed programming languages, such as Python and JavaScript, have become extremely popular, and large portions of newly written code are in one of these languages.
While the lack of static type annotations enables fast prototyping, it often leads to problems when projects grow.
Examples include type errors that remain unnoticed for a long time~\cite{Gao17}, suboptimal IDE support, and difficult to understand APIs~\cite{Mayer12}.
To solve these problems, in recent years, many dynamic languages obtained support for \emph{type annotations}, which enable programmers to specify types in a fashion similar to a statically typed language.
Type annotations are usually ignored at runtime; nevertheless, they serve both as hints for developers using external APIs and as inputs to gradual type checkers that ensure that specific programming errors cannot occur.
To cope with legacy code bases, type annotations can be introduced gradually; in such cases, the type checker will check only code that is annotated.

As manually annotating code is time-consuming and error-prone~\cite{Ore2018}, developers must resort to automated methods.
One way to address the lack of type annotations is type inference via traditional static analysis.
Unfortunately, dynamic features, such as heterogeneous arrays, polymorphic variables, dynamic code evaluation, and monkey patching make static type inference a hard problem for popular dynamic languages, such as Python or JavaScript~\cite{Chandra2016}.
Static type inference tools typically handle these challenges by inferring a type only if it is certain or very likely (under some assumptions), which significantly limits the number of types that can be inferred.

Motivated by the inherent difficulties of giving definitive answers via static analysis, several probabilistic techniques for predicting types have been proposed.
A popular direction is to exploit the existence of already annotated code as training data to train machine learning models that then predict types in not yet annotated code.
Several approaches predict the type of a code entity, e.g., a variable or a function, from the code contexts in which this entity occurs~\cite{Raychev2015,Hellendoorn2018}.
Other approaches exploit natural language information embedded in source code, e.g., variable names or comments, as a valuable source of informal type hints~\cite{Malik2019,Xu2016}.

While existing approaches for predicting types are effective in some scenarios, they suffer from \emph{imprecision} and \emph{combinatorial explosion}.
Probabilistic type predictors are inherently imprecise because they suggest one or more likely types for each missing annotation, but cannot guarantee their correctness.
The task of deciding which of these suggestions are correct is left to the developer.
Because probabilistic predictors suggest a ranked list of likely types, choosing a type-correct combination of type annotations across multiple program elements causes combinatorial explosion.
A na\"ive approach would be to let a developer or a tool choose from all combinations of the predicted types.
Unfortunately, this approach does not scale to larger code examples, because the number of type combinations to consider is exponential in the number of not yet annotated code entities.

%	\item \emph{Variety of type hints}.
%The best existing probabilistic type prediction approaches predict types based on either code context or natural language information, but not both.
%This limitation prevents probabilistic type prediction from unleashing the full potential of learning from already annotated code.
%A na\"ive solution would be to query multiple predictors and to combine their type suggestions, e.g., by always giving priority to one tool and consulting the others only if that tool gives no prediction at all~\cite{Hellendoorn2018}.
%However, this approach enforces a strict ordering of predictors and misses opportunities to reconcile multiple kinds of information into a single prediction.
%\end{itemize}

This paper presents \name{}, a combination of learning-based, probabilistic type prediction and a feedback-directed, search-based validation of predicted types.
The approach addresses the imprecision problem based on the insight that a gradual type checker can pinpoint contradictory type annotations, which guides the selection of suitable types from the set of predicted types.
To make the search for a consistent set of types tractable, we formulate the problem as a combinatorial search and present a search strategy that finds type-correct type annotations efficiently.
\name{} makes use of the variety of type hints present in code through a novel neural architecture that exploits both natural language, in the form of identifier names and code comments, similar to prior work~\cite{Malik2019}, and also programming context, in the form of usage sequences.

\begin{figure}
	\begin{lstlisting}
# Predicted argument type: int, str, bool
# Predicted return type: str, Optional[str], None
def find_match(color):
  """
  Args:
    color (str): color to match on and return
  """
  candidates = get_colors()
  for candidate in candidates:
    if color == candidate:
      return color
  return None

# Predicted return types: List[str], List[Any], str
def get_colors():
  return ["red", "blue", "green"]
	\end{lstlisting}
	\caption{Example of search for type-correct predicted types.}
	\label{fig:search-example}
\end{figure}

To illustrate the approach, consider the two to-be-typed functions in Figure~\ref{fig:search-example}.
Given this code, the neural type model of \name{} predicts a ranked list of likely types for each argument type and return type, as indicated by the comments.
\name{} starts by adding the top-ranked predictions as type annotations, which introduces a type error about an incorrect return type of \code{find\_match}, though.
Based on this feedback, the search tries to change the return type of \code{find\_match} to the second-best suggestion, \code{Optional[str]}.
Unfortunately, this combination of added types leads to another type error because the return type is inconsistent with the argument \code{key} being of type \code{int}.
The search again refines the type annotations by trying to use the second-best suggestion, \code{str}, for the argument \code{key}.
Because the resulting set of type annotations is type-correct according to the type checker, \name{} adds these types to the code.

%1) Search initially adds all top-1 predictions as annotations.
%- Type warning about return type of find_match
%2) Search tries to swap return type of find_match to Optional[str]
%- Different type warning about return type of find_match
%3) Search tries to swap type of arg key to Str
%- Success!

We implement \name{} for Python and apply it on two large code bases: a multi-million line code base at \fb{} that powers applications used by billions of people, and a corpus of popular open-source projects.
We show that the neural model predicts individual types with a precision of 64\% (85\%, 91\%) and a recall of 52\% (64\%, 68\%) within the top-1 (top-3, top-5) predictions, which outperforms a recent, closely related approach~\cite{Malik2019} by 10\% and 6\% respectively.
Based on this model, the feedback-directed search finds a type-correct subset of type annotations that can produce complete and type-correct annotations for 42\% to 64\% of all files.
Comparing \name{} with a traditional, static type inference shows that both techniques complement each other and that \name{} predicts many more types than traditional type inference.
In summary, this paper makes the following contributions:
\begin{itemize}
	\item A combination of probabilistic type prediction and search-based validation of predicted types.
	The feedback-directed search for type-correct types can be used with any probabilistic type predictor and any gradual type checker.
	\item A novel neural type prediction model that exploits both code context and natural language information.
	\item Empirical evidence that the approach is effective for type-annotating large-scale code bases with minimal human effort.
	The initial experience from using \name{} at \fb{} on a code base that powers tools used by billions of people has been positive.
\end{itemize}

\section{Approach}

\begin{figure*}
	\includegraphics[width=.9\linewidth]{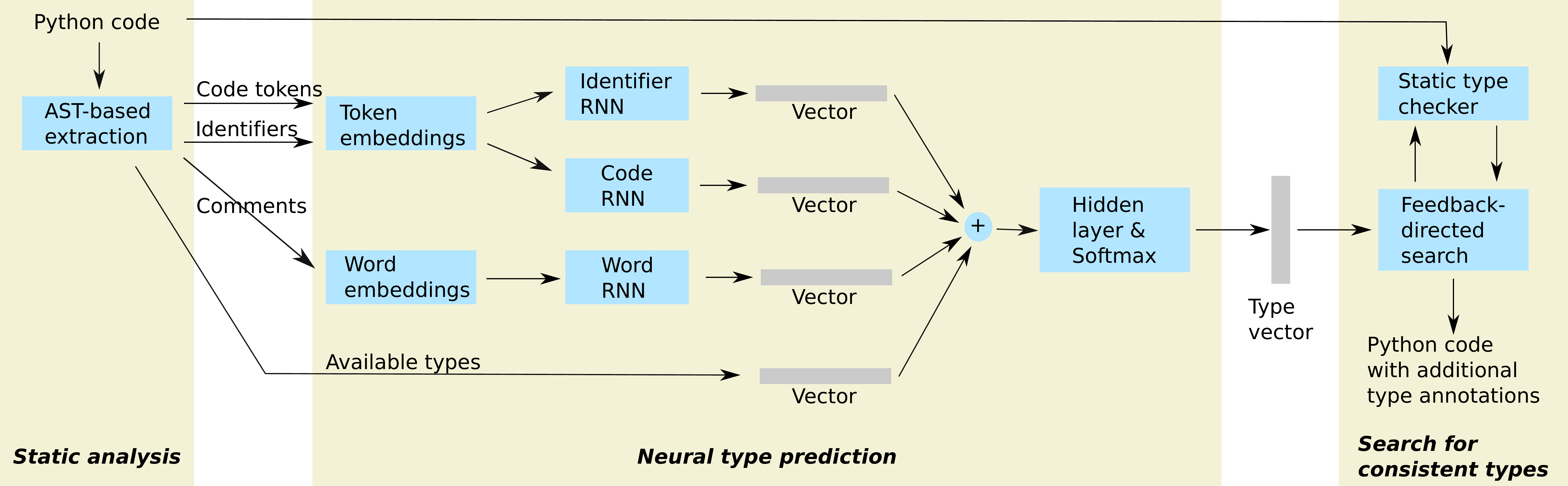}
	\caption{Overview of \name{}.}
	\label{fig:overview}
\end{figure*}

Figure~\ref{fig:overview} gives a high-level overview of the \name{} approach.
The input to \name{} is a corpus of code where some, but not all types are annotated.
The approach consists of three main parts.
First, a lightweight static analysis extracts several kinds of information from the given code (Section~\ref{sec:static analysis}).
The extracted information includes programming structure information, such as usages of a function's arguments, and natural language information, such as identifier names and comments.
Next, a neural type predictor learns from the already annotated types and their associated information how to predict missing types (Section~\ref{sec:typepred}).
Once trained, this model can predict likely types for currently unannotated parts of the code.
Finally, a feedback-directed search uses the trained model to find a type assignment that is consistent and type-correct according to a static, gradual type checker (Section~\ref{sec:search}).
The overall output of \name{} is code with additional type annotations.

\subsection{Static Extraction of Types and Context Information}
\label{sec:static analysis}

The first part of \name{} is an AST-based static analysis that extracts types and context information useful to predict types.
The analysis is designed to be lightweight and easy to apply to other programming languages.
We currently focus on function-level types, i.e., argument types and return types.
These types are particularly important for two reasons:
(i) Given function-level types, gradual type checkers can type-check the function bodies by inferring the types of (some) local variables.
(ii) Function-level types serve as interface documentation.

For each type, the static analysis gathers four kinds of context information, which the following describes and illustrates with the example in Figure~\ref{fig:example}.
Each of the four kinds of information may provide hints about an argument type or return type, and our model learns to combine these hints into a prediction of the most likely types.

\begin{figure}
	\begin{lstlisting}
from html import HtmlElement

def update_name(name, do_propagate, element):
  """ Update the name and (optionally)
      propagate to dependents. """
  first_name = name.split(" ")[0] /*#\label{line:name}#*/
  element.first = first_name
  if do_propagate:
    for d in dependents:
      d.notify(NAME, first_name)
	\end{lstlisting}
	\caption{Example of a to-be-typed Python function.}
	\label{fig:example}
\end{figure}

\paragraph{Identifier names associated with the to-be-typed program element}
As shown by prior work~\cite{Malik2019,oopsla2018-DeepBugs}, natural language information embedded in source code can provide valuable hints about program properties.
For example, the argument names \code{name} and \code{do\_pro\-pa\-ga\-te} in Figure~\ref{fig:example} suggest that the arguments may be a string and a boolean, respectively.
To enable \name{} to benefit from such hints, the static analysis extracts the identifier name of each with function and each function argument.

\paragraph{Code occurrences of the to-be-typed program element}
In addition to the above natural language information, \name{} exploits programming language type hints.
One of them is the way a to-be-typed program element is used:
As a hint about argument types, the analysis considers all usages of an argument within the function body.
Another kind of information is code that defines the to-be-typed program element:
As a hint about return types, the analysis considers all return statements in a function.
For each of these code locations, the analysis extracts the corresponding sequence of code tokens $o_1,...,o_k$.
Specifically, the analysis extracts a window of tokens around each occurrence of an argument (default size of window: 7) and all tokens of a return statement.
For example, the analysis extracts the token sequence \code{$\backslash$n}, \code{first\_name}, \code{=}, \code{name}, \code{.}, \code{split}, \code{(} around the usage of \code{name} at line~\ref{line:name}.

As an alternative to extracting token sequences, \name{} could perform a more sophisticated static analysis, e.g., by tracking data flows starting at arguments or ending in return values.
We instead focus on token sequences because it provides a sufficiently strong signal, scales well to large code bases, and could be easily ported to other programming languages.

\paragraph{Function-level comments}
Similar to identifier names, comments are another informal source of hints about types.
For the example in Figure~\ref{fig:example}, a developer might infer from the function-level comment that the function has some side effects but probably does not return any value.
To allow the approach to benefit from such hints, the static analysis extracts all function-level comments, i.e., docstrings in Python.
For a given function, the approach uses this comment both for predicting the argument types and the return type of the function.

\paragraph{Available types}
To annotate a type beyond the built-in types of Python, the type needs to be either imported or locally defined.
Because types used in an annotation are likely to be already imported or locally defined, the analysis extracts all types available in a file.
To this end, the analysis parses all \code{import} statements and all class definitions in a file.
For the example in Figure~\ref{fig:example}, the analysis will extract \code{HtmlElement} as an available type, which hints at the argument \code{element} being of this type.

\medskip
\noindent
Based on these four kinds of type hints, the analysis extracts the following information for argument types and return types, respectively:

\begin{definition}[Argument type information]
\label{def:argTypeInfo}
	For a function argument $a$, the statically extracted information is a tuple
	$$(n_{\mathit{fct}}, n_{\mathit{arg}}, N_{\mathit{args}}, c, U, t)$$ where
    $n_{\mathit{fct}}$ is the function name,
	$n_{\mathit{arg}}$ is the name of the argument $a$,
	$N_{\mathit{args}}$ is the sequence of names of other arguments (if any),
	$c$ is the comment associated with the function,
	$U$ is a set of usage sequences, each of which is a sequence $o_1, ..., o_k$ of tokens, and
	$t$ is the type of the argument.
\end{definition}

\begin{definition}[Return type information]
\label{def:retTypeInfo}
	For the return type of a function $f$, the statically extracted information is a tuple
	$$(n_{\mathit{fct}}, N_{\mathit{args}}, c, R, t)$$ where
	$n_{\mathit{fct}}$ is the function name,
	$N_{\mathit{args}}$ is the sequence of argument names,
	$c$ is the comment associated with the function,
	$R$ is a set of return statements, each of which is a sequence $o_1, ..., o_k$ of tokens, and
	$t$ is the return type of $f$.
\end{definition}

If any of the above information is missing, the corresponding elements of the tuple is filled with a placeholder.
In particular, the static analysis extracts the above also for unannotated types, to enable \name{} to predict types based on the context.

\subsection{Neural Type Prediction Model}
\label{sec:typepred}

Given the extracted types and context information, the next part of \name{} is a neural model that predicts the former from the latter.
We formulate the type prediction problem as a classification problem, where the model predicts a probability distribution over a fixed set of types.
The neural type prediction model, summarized in the middle part of Figure~\ref{fig:overview}, combines the four kinds of information described in Section~\ref{sec:static analysis} into a single type prediction.

To represent identifier names, source code tokens, and words in a way suitable for learning, \name{} maps each into a real-valued vector using a Word2Vec~\cite{Mikolov2013a} embedding.
We train two embeddings, a \emph{code embedding} $E_{code}$ for code tokens and identifier names, and a \emph{word embedding} $E_{word}$ for words in comments.
$E_{code}$ is trained on sequences of tokens extracted from source code files,
while $E_{word}$ is trained on sequences of words extracted from comments.
To mitigate the problem of large vocabularies in source code~\cite{Babii2019}, \name{} preprocesses each identifier using a helper function $\mathit{norm()}$, which tokenizes, lemmatizes, and lowercases each identifier.

\subsubsection{Learning from Identifiers}

This neural submodel learns from the identifier names of functions and function arguments.
The model combines all identifiers associated with a type into sequence.
Given argument type information $(n_{\mathit{fct}}, n_{\mathit{arg}},\allowbreak N_{\mathit{args}}, c, U, t)$, the sequence is
$$\mathit{norm}(n_{\mathit{arg}}) \circ s \circ \mathit{norm}(n_{\mathit{fct}}) \circ \mathit{norm}(N_{\mathit{args}})$$
where $\circ$ flattens and concatenates sequences, and $s$ is a separator.
Given return type information $(n_{\mathit{fct}}, N_{\mathit{args}}, c, R, t)$, the sequence is
$$\mathit{norm}(n_{\mathit{fct}}) \circ s \circ \mathit{norm}(N_{\mathit{args}})$$
For example, the sequence for the return type of the function in Figure~\ref{fig:example} is ``update name $s$ name do propagate element''.

\name{} learns from these sequences of words by summarizing them into a single vector using a bi-directional, recurrent neural network (RNN) based on LSTM cells.
To ease parallelization, we pad sequences that are too short and truncate sequences that are too long (default length: 10).
The final hidden states of the RNN serve as a condensed vector representation, $v_{\mathit{ids}}$, of all identifier-related hints.

\subsubsection{Learning from Token Sequences}

This neural submodel learns from source code information associated with a type.
Similar to the submodel for identifiers, this submodel composes all relevant tokens into a sequence and summarize them into a single vector $v_{\mathit{code}}$ using an RNN.
For arguments and return types, the sequence consists of the tokens involved in the usages $U$ (Definition~\ref{def:argTypeInfo}) and the return statements $R$ (Definition~\ref{def:retTypeInfo}), respectively.
Before feeding these sequences into an RNN, we bound the length of each token sequence (default: $k=7$) and of the number of token sequences (default: 3).

\subsubsection{Learning from Comments}

This neural submodel learns type hints from comments associated with a function.
To this end, \name{} splits a given comment into a sequence of words, bound the length of the sequence to a fixed value (default: 20), and summarizes the sequence via another RNN.
The result is a fixed-length vector $v_{\mathit{comments}}$.

\subsubsection{Learning from Available Types}

The fourth kind of information that \name{} learns from is the set of types available in the current source code file.
The approach assumes a fixed-size vocabulary $T$ of types (default size: 1,000).
This vocabulary covers the vast majority of all type occurrences because most type annotations either use one of the built-in primitive types, e.g., \code{str} or \code{bool}, common non-primitive types, e.g., \code{List} or \code{Dict}, or their combinations, e.g., \code{List[int]} or \code{Dict[str, bool]}.
Any types beyond the type vocabulary are represented as a special ``unknown'' type.

To represent which types are available, we use a binary vector of size $T$, called the \emph{type mask}.
Each element in this vector represents one type, and an element is set to one if and and only if its type is present.
The resulting vector $v_{\mathit{availTypes}}$ of available types is passed as-is into the final part of the neural model.

\subsubsection{Predicting the Most Likely Type}

The final submodel concatenates the four vectors $v_{\mathit{ids}}$, $v_{\mathit{code}}$, $v_{\mathit{comments}}$, and $v_{\mathit{availTypes}}$ into a single vector and passes it through a fully connected layer that predicts the most likely type.
The output vector has size $|T|$ and represents a probability distribution over the set of types.
For example, suppose the type vocabulary had only four types \code{int}, \code{bool}, \code{None}, and \code{List}, and that the output vector is $[0.1, 0.6, 0.2, 0.1]$.
In this case, the model would predict that \code{bool} is the most likely type, following by \code{None}.

There are two ways to handle uncertainty and limited knowledge in the model.
First, we interpret the predicted probability of a type as a confidence measure and only suggest types to a user that are predicted with a confidence above some configurable threshold.
Second, we encode types not included in the fixed-size type vocabulary as a special ``unknown'' type.
The model hence learns to predict ``unknown'' whenever none of the types in the vocabulary fit the given context information.
During prediction, \name{} never suggests the ``unknown'' type to the user, but instead does not make any suggestion in case the model predicts ``unknown''.

\subsubsection{Training}

To train the type prediction model, \name{} relies on already type-annotated code.
Given such code, the approach creates one pair of context information and type for each argument type and for each return type.
These pairs then serve as training data to tune the parameters of the different neural submodels.
We use stochastic gradient descent, the Adam optimizer, and cross-entropy as the loss function.
The entire neural model is learned jointly, enabling the model to summarize each kind of type hint into the most suitable form and to decide which type hints to consider for a given query.
We train two separate models for argument types and function types, each learned from training data consisting of only one kind of type.
The rationale is that some of the available type hints need to be interpreted differently depending on whether the goal is to predict an argument type or a return type.

\subsection{Feedback-guided Search for Consistent Types}
\label{sec:search}

\begin{algorithm}
  \caption{Find a correct type assignment for a file $f$}
  \label{twsearch}
  \begin{algorithmic}[1]
    \Function{assign\_types}{$f$}
      \State $T \gets $ all type slots in $f$
      \State $\mathcal{P}_t^{1..k} \gets \{\text{predictions}(t, k) \mid t \in T\}$ \Comment{Top $k$ predictions}
      \State $a \gets \{\mathcal{P}_t^1 \mid t \in T\}$ \Comment{Initial type assignment}
      \State $a.\text{score} \gets \text{typecheck}(a, f)$ \Comment{Feedback
       function}

      \State $\text{work\_set} \gets \text{new\_states}(a, \mathcal{P}, T)$
      \State $\text{done} \gets \{a\}$
      \While{$\text{min}(\{x.\text{score} \mid x \in \text{done}\}) > 0 \wedge \text{work\_set} \neq \emptyset$}
	      \State $a \gets \text{pick}(\text{work\_set})$ \Comment{Biased random selection}
	      \State $a.\text{score} \gets \text{typecheck}(a, f)$

	 	  \If{$\text{greedy} \wedge a.\text{score} < a.\text{parent}.\text{score}$}
	  	      \State $\text{work\_set} \gets \text{new\_states}(a, \mathcal{P}, T)$
	  	  \ElsIf{$\text{non-greedy}$}
	  	  	  \State $\text{work\_set} \gets \text{work\_set} \cup
	  	  	  		      (\text{new\_states}(a, \mathcal{P}, T) \setminus \text{done}) $
	 	  \EndIf
	      \State $\text{done} \gets \text{done} \cup \{ a \}$
      \EndWhile
      \State \Return{$\text{argmin}(\{x.\text{score} \mid x \in \text{done}\})$}
  \EndFunction

    \Function{new\_states}{$a, \mathcal{P}, T$}
	    \State children $\gets \{\}$
	    \ForAll{$t \in T$}
	    	\ForAll{$\mathcal{P}^j_t$ where $j>$ rank of current $a[t]$}
	    		\State $a_{child} \gets$ modify $a$ to use $\mathcal{P}^j_t$ at $t$
	    		\State children $\gets \{ a_{child} \}$
	    	\EndFor
	    	\State $a_{child} \gets$ modify $a$ to not use any type at $t$
	    	\State children $\gets \{ a_{child} \}$
	    \EndFor
		\State \Return children
    \EndFunction
  \end{algorithmic}

\end{algorithm}

The neural type prediction model provides a ranked list of $k$ predictions for each missing type annotation.
Given a set of locations for which a type annotation is missing, called \emph{type slots}, and a list of probabilistic predictions for each slot, the question is which of the suggested types to assign to the slots.
A na\"ive approach might fill each slot with the top-ranked type.
%Instead of letting developers struggle with this question, \name{} searches for a subset of all predicted types, so that all types in the subset are consistent with each other and do not introduce any type errors.
%The following presents our approach for performing this search.
%We argue that, in principle, this validation step must be part of any probabilistic type predictor, including previously proposed machine learning models~\cite{Raychev2015,Hellendoorn2018,Malik2019} and other probabilistic approaches~\cite{Xu2016}.
%
%For some partially annotated code, we call a missing type annotation a \emph{type slot}, which needs to be filled by a type, and we call a mapping of type slots to types a \emph{type assignment}.
%Given the type suggestions predicted by the model for each type slot, a na\"ive approach would be to fill each slot with the top-ranked type.
However, because the neural model may mis-predict some types, this approach may yield type assignments where the added annotations are not consistent with each other or with the remaining program.
%For example, suppose that the type prediction model suggests type \code{int} for the argument \code{name} in Figure~\ref{fig:example}.
%Adding this annotation would be inconsistent with the way \code{name} is used in the function body, where the argument is assumed to be a string.

To avoid introducing type errors, \name{} leverages an existing gradual type checker as a filter to validate candidate type assignments.
Such type checkers exist for all popular dynamically typed languages that support optional type annotations, e.g., \tool{pyre} and \tool{mypy} for Python, and \tool{flow} for JavaScript.
%For our running example, incorrectly annotating \code{name} with type \code{int} will yield a type error at line~\ref{line:name} because \code{int} values do not provide a \code{split} function.
\name{} exploits feedback from the type checker to guide a search for consistent
types, as presented in \Cref{twsearch} and explained in the sections below.
%The following describes the search space \name{} explores (Section~\ref{sec:search space}), how to encode the type checker feedback into a numeric feedback function (Section~\ref{sec:feedback function}), and how \name{} explores the search space based on the feedback (Section~\ref{sec:search strategies}).

\subsubsection{Search Space}
\label{sec:search space}

Given a set~$T$ of type slots and $k$ predicted types for each slot, we formulate the problem of finding a consistent type assignment as a combinatorial search problem.
The search space consists of the set $\mathcal{P}$ of possible \emph{type assignments}.
%This set has $2^{|S|}$ elements, i.e., the search space is exponential in the number of missing types.
%For the example in Figure~\ref{fig:example}, there are four type slots: three for the argument types and one for the return type of the function.
%Because the given code does not come with any type annotations, the starting point of the search space is an empty type assignment.
%From this starting point, there are $2^4=16$ possible type assignments: use all predicted types, none of the predicted types, or any subset of them.
%An alternative formulation of the search problem could consider multiple type predictions for each type slot.
For $|T|$ type slots and $k$ possible types for each slot, there are $(k+1)^{|T|}$ type assignments (the $+1$ is for not assigning any of the predicted types).
%Both formulations of the problem are exponential  in the number of missing types.
%We here focus on the first formulation, i.e., only one possible type per slot, because it reduces the search space and because we find the top-most prediction to often be correct.

\subsubsection{Feedback Function}
\label{sec:feedback function}

%The search procedure aims to find a type assignment that minimizes the number of unfilled type slots and precludes type errors from being introduced by the assignment.
Exhaustively exploring $\mathcal{P}$ is practically infeasible for files with many missing types, because invoking the gradual type checker is relatively expensive (typically, in the order of several seconds per file).
Instead, \name{} uses a feedback function (\textsf{typecheck}) to efficiently steer toward the most promising type assignments.

The feedback function is based on two values, both of which the search wants to minimize:
\begin{itemize}
	\item $n_{\mathit{missing}}$ : The number of missing types.
	\item $n_{\mathit{errors}}$ : The number  of type errors.
\end{itemize}
\name{} combines these into a weighted sum $\mathit{score} = v \cdot n_{\mathit{missing}} + w \cdot n_{\mathit{errors}}$.
By default, we set $v$ to $1$ and $w$ to the number of initially missing types plus one, which is motivated by the fact that adding an incorrect type often leads to an additional error.
By giving type errors a high enough weight, we ensure that the search never returns a type assignment that adds type errors to the code.

%A particular challenge in measuring $n_{\mathit{errors}}$ is that adding a type annotation to a function may cause type errors, irrespective of what type gets added.
%Since many gradual type checkers only type-check a function body if the function has at least its return type annotated, it may be the case that adding any type,
%including the correct type, as the return type may yield additional type errors.
%For example, consider the following function:
%\begin{lstlisting}[numbers=none]
%def has_name(name):
%    s = 23
%    s = "a"  # type error: incompatible types 'str' and 'int'
%    return name == s
%\end{lstlisting}
%The type error caused by assigning both a string and an integer to the same variable is unrelated to argument types and return types that \name{} aims to add.
%However, since the error is reported only once a return type annotation gets added, it would confuse the search to believe that the error is caused by the added annotation.
%%
%As such pre-existing type errors do not provide feedback about whether a newly added type is correct, we ignore them in $n_{\mathit{errors}}$.
%% To this end, \name{} initially adds a certainly wrong return type to functions featuring return type slots, checks which type errors appear in the function bodies, and then filters these errors when evaluating other type assignments.
%Before beginning to search, \name{} mines these errors by initially adding a certainly wrong return type to functions with missing return types and checking which type errors appear in the function bodies.

\subsubsection{Exploring the Search Space}
\label{sec:search strategies}

\name{} explores the space of type assignments through an optimistic search strategy (Algorithm~\ref{twsearch}).
It assumes that most predictions are correct, and then refines type annotations to minimize the feedback function.
Each exploration step explores a state $a$, which consists of a type
assignment, the score computed by the feedback function, and a link to the parent state.
The initial state is generated by retrieving the top-1 predictions
from $\mathcal{P}$ for each type slot $t$ and invoking the feedback
function (lines~4 and~5).
The next states to be explored are added to a work set, while the explored
states are kept in the ``done'' set.
The algorithm loops over items in the work set until either the feedback score has been
minimized or the search explored all potential type assignments (line 8).
The assignment with the minimal score is returned as a result (line 18).

To retrieve the next type assignments to possibly explore from the current state, \name{} invokes the \textsf{new\_states} helper function.
It adds all type assignments that can be obtained from the current state by modifying exactly one type slot, either by using a lower-ranked type suggestion or by not adding any type for this slot (lines~22 to~29).

The main loop of the algorithm (lines~8 to~17) picks a next state to evaluate from the working set (line ~9), queries the feedback function (line~10) and updates the done set
with the explored state (line~16).
The \textsf{pick} function is a biased random selection that prefers states based on two criteria.
First, it prefers states that add more type annotations over states that add fewer annotations.
Second, it prefers states that modify a type annotation at a line close to a line with a type error.
Intuitively, such states are more likely to fix the cause of a type error than a randomly selected state.\footnote{The reason for relying on line numbers as the interface between
the type checker and \name{} is to enable plugging any type checker into our
search.}
The working set is updated with all new states that have not been currently
explored.

\name{} implements two variants of the search, a greedy and a non-greedy one.
The \emph{greedy} strategy aggressively explores children of type assignment
that decrease the feedback score and prunes children of states that increase it
(line 12).
% The non-greedy strategy first explores the siblings of children that increase the search score before pruning the whole branch; it then selectively prunes only those siblings that increase the score.
The \emph{non-greedy} performs no pruning, i.e., it can explore a larger part of
the search space at the expense of time (line 14).

As an optimization of Algorithm~\ref{twsearch}, \name{} invokes the \textsf{assign\_types} function twice.
The first call considers only type slots for return types, whereas the second call considers all type slots for argument types.
The reason for this two-phase approach is that many gradual type checkers, including pyre, the one used in our evaluation, type-check a function only if its return type is annotated.
If \name{} would add argument type annotations before adding return type annotations, the feedback function might not include all type errors triggered by an incorrectly added argument annotation.

\section{Implementation}

The implementation of \name{} builds upon a variety of tools in the Python ecosystem.
For the static analysis phase, we apply a data extraction pipeline consisting of Python's own \textsf{ast} library to parse the code into an AST format,
and NLTK and its WordNetLemmatizer module to perform standard NLP tasks (lemmatization, stop word removal).
The pipeline is parallelized so that it handles multiple files concurrently.
The neural network model is implemented in PyTorch.
For obtaining embeddings for words and tokens, we pre-train a Word2Vec model using the \textsf{gensim} library.
The search phase of \name{} builds upon the LibCST\footnote{https://github.com/Instagram/LibCST} to add types to existing Python files.
We use \tool{pyre} for static type checking.
Our LSTM models all use 200-dimensional hidden layers, and we train for 10 epochs with a learning rate of 0.005 using the Adam Optimizer.
%
%All experiments where run on a server featuring 48 Xeon E5-2680 2.5GHz CPUs, 256GB RAM
%and 2 NVidia Tesla M40 GPUs with 12GB RAM each.

\section{Evaluation}

We structure our evaluation along four research questions.

\noindent{\bfseries{RQ~1}}: How effective is \name{}'s model at predicting argument and return types, and how does it compare to existing work?

\noindent{\bfseries{RQ~2}}: How much do the different kinds of context information contribute to the model's prediction abilities?

\noindent{\bfseries{RQ~3}}: How effective is \name{}'s search?

\noindent{\bfseries{RQ~4}}: How does \name{} compare to traditional static type inference?

%\vspace{0.5em}
%\noindent{\bfseries{RQ~5}}: How efficient is \name{}?
\subsection{Datasets}

\begin{table}
	\caption{Internal and open-source datasets.}
	\label{tab:datasets}
	\begin{small}
	\begin{tabular}{@{}lrr@{}}
		\toprule
		Metric & Internal & OSS \\
		\midrule
		Repositories & 1 & 1,137 \\
		Files &
		$ * $ %170,470
		& 11,993
		\\
		Lines of code &
		$ * $ %39M
		& 2.7M \\
		\midrule
		\textsf{Functions} &
		$ * $ %765,725
		& 146,106 \\
		\dots with return type annotation &
		68\% % 68,070
		& 80,341 (55\%)\\
		\dots with comment &
		21.8\% % 157,570
		& 53,500 (39.3\%)  \\
		\dots with both &
		16\% % 17,637
		& 32,409 (22.2\%) \\
		\dots ignored because trivial &
		7.4\% % 60,248
		& 12,436 (8.5\%) \\
		\midrule
		\textsf{Arguments} &
		* %1,513,221
		&  274,425 \\
		\dots with type annotation &
		50\% %98,023
		& 112,409 (41\%) \\
		\dots ignored because trivial &
		33\% %521,916
		& 96,036 (35\%) \\
		\midrule
		\textsf{Types - Return} & * & 7,383 \\
		\dots occurrences ignored (out of vocab.) & 20.2\% & 11.3\% \\
		\midrule
		\textsf{Types - Argument} & * & 8,215 \\
		\dots occurrences ignored (out of vocab.) & 21.3\% & 13.7\% \\
		\midrule
		\textsf{Training time (min:sec)} & & \\
		\dots parsing &
		several minutes % 8:05 + 4:35
		& 1:45 \\
		\dots training embeddings &
		several minutes % 0:47 + 14:41
		& 2:29 \\
		\dots training neural model &
		several minutes %12:45
		& 2:20 \\
		\bottomrule
		\multicolumn{3}{l}{$*$ = not available for disclosure}
	\end{tabular}
	\end{small}
\end{table}

\name{} is developed and evaluated within \fb{}.
As the internal code base is not publicly available and to ensure that the presented results are replicable, we use two datasets:

\begin{description}

\item[Internal code base] We collect Python from a large internal code repository.
\item[OSS corpus] We search GitHub for all projects tagged as \texttt{python3}.
We also search Libraries.io for all Python projects that include \texttt{mypy} as a dependency.
We then remove all projects that have less than 50 stars on GitHub, to ensure that the included projects are of substantial public interest.
To ease future work to compare with \name{}, all results for the OSS corpus are available for download.\footnote{\url{http://software-lab.org/projects/TypeWriter/data.tar.gz}}

\end{description}

The resulting dataset statistics can be found in~\Cref{tab:datasets}.
The internal dataset is much larger in size, but both datasets are comparable in terms of the percentage of annotated code.
By restricting the type vocabulary to a fixed size, we exclude around 10\% of all type occurrences for both datasets.
This percentage is similar for both datasets, despite their different sizes, because types follow a long-tail distribution, i.e., relatively few types account for the majority of all type occurrences.
We ignore some types because they are \emph{trivial} to predict, such as the return type of \code{\_\_str\_\_}, which always is \code{str}, or the type of the \code{self} argument of a method, which always is the surrounding class.
\name{} could easily predict many of these trivial types, but a simple syntactic analysis would also be sufficient.
We ignore trivial types for the evaluation to avoid skewing the results in favor of \name{}.

\subsection{Examples}

\begin{figure}
	\begin{lstlisting}
  # PrefectHQ/ct/f/blob/master/src/prefect/utilities/notifications.py
  # Commit: 864d44b
  # Successful annotation of return type
  def callback_factory(...) -> Callable:
    """
    ...
    Returns:
        - state_handler (Callable): a state handler function that
        can be attached to both Tasks and Flows
    ...
    """
    def state_handler(...):
        ...
    return state_handler
	\end{lstlisting}
	% source : https://github.com/PrefectHQ/prefect/blob/master/src/prefect/cli/auth.py#L35
	Example 1
%	\vspace{1em}
%
	% \begin{lstlisting}
	% # __main__.py
	% # Correct annotation of argument 'loop'
	% def exit_from_event_loop_thread(
	%   loop: asyncio.AbstractEventLoop, stop
	% ) -> None:
	%   loop.stop()
	%   if not stop.done():
	%     # When exiting the thread that runs the event loop, raise
	%     # KeyboardInterrupt in the main thead to exit the program.
	%     try:
	%       ctrl_c = signal.CTRL_C_EVENT  # Windows
	%     except AttributeError:
	%       ctrl_c = signal.SIGINT  # POSIX
	%     os.kill(os.getpid(), ctrl_c)
	% \end{lstlisting}
	% Example 1
	% https://github.com/aaugustin/websockets/blob/master/src/websockets/__main__.py

	\vspace{0.5em}

%	\begin{lstlisting}
%	# Incorrect annotation of argument key
%	# (Expected Optional[str])
%	def set(self, key: str, value):
%	  """Set value."""
%	  _set = self._map.get(key)
%	  if not _set:
%	    _set = list()
%	    self._map[key] = _set
%	  ...
%	\end{lstlisting}
%	Example 3
%	% https://github.com/miyakogi/pyppeteer/blob/dev/pyppeteer/multimap.py
%
%	\vspace{1em}
%
	\begin{lstlisting}
  # awslabs/sockeye/blob/master/sockeye/average.py
  # Commit: bcda569
  # Incorrect annotation of return type: expected List[str]
    def find_checkpoints(...) -> List[Path]:
        """
        ...
        :return: List of paths corresponding to chosen checkpoints.
        """
        ...
        params_paths = [
            os.path.join(model_path, C.PARAMS_NAME % point[-1])
            for point in top_n
        ]
        ...
        return params_paths
	\end{lstlisting}
	% https://github.com/aaugustin/websockets/blob/master/src/websockets/headers.py
	Example 2

	\caption{Examples of successful and unsuccessful type predictions (GitHub: \texttt{PrefectHQ/ct}, \texttt{awslabs/sockeye}).}
	\label{fig:eval_examples}
\end{figure}

Figure~\ref{fig:eval_examples} shows examples of successful and unsuccessful type predictions in the OSS dataset.
%Example 1 shows an example of TypeWriter successfully adding a new annotation that we manually evaluate to be correct.
%In this instance, from both the docstrings and our understanding of authentication tokens, we agree that the most likely type of token is a string.
Example 1 presents a case where TypeWriter correctly predicts a type annotation.
Here, the code context and comments provide enough hints indicating that token is of type \texttt{Callable}.
Example 2 presents a case where \name{} does not correctly predict the type, but the prediction is close to what is expected.
We hypothesize that this case of mis-prediction is due to the fact that \name{} tries to associate associations between natural language and types, or in this case, the word ``path'' and the type \texttt{Path}.

\subsection{RQ~1: Effectiveness of the Neural Model}
\label{sec:effectnn}
\paragraph{Prediction tasks}
To evaluate the neural type prediction, we define two prediction tasks:
(i) \textsf{ReturnPrediction}, where the model predicts the return types of
functions, and
(ii) \textsf{ArgumentPrediction}, where the model predicts the types of function arguments, and
%(iii) \textsf{CombinedPrediction}, where the model predicts both return and argument types.

\begin{table*}
	\vspace{-1em}
	\caption{Effectiveness of neural type prediction.}
	\label{tab:effectiveness_model}
	\vspace{-1em}

	\begin{small}
	\setlength{\tabcolsep}{8pt}
	\begin{tabular}{lllrrrrrrrrrr}
		\toprule
		Corpus & Task & Model &
		\multicolumn{3}{c}{Precision} &
		\multicolumn{3}{c}{Recall} &
		\multicolumn{3}{c}{F1-score} \\
		\cmidrule{4-6}
		\cmidrule{7-9}
		\cmidrule{10-12}
		&&& Top-1 & Top-3 & Top-5 &
		Top-1 & Top-3 & Top-5 &
		Top-1 & Top-3 & Top-5 \\
		\midrule
		\midrule
		\textsf{Internal} & \textsf{ReturnPrediction} & \name{} & 0.73  & 0.88 & 0.92 & 0.58 & 0.66 & 0.69 & 0.64 & 0.76 & 0.79\\
		 &  & NL2Type & 0.60  & 0.82 & 0.88 & 0.50 & 0.61 & 0.65 & 0.55 & 0.70 & 0.75\\
		 &  & DeepTyper & 0.70  & 0.87 & 0.92 & 0.43 & 0.54 & 0.59 & 0.53 & 0.67 & 0.72\\
		 &  & Na\"ive baseline & 0.15  & 0.22 & 0.25 & 0.24 & 0.40 & 0.45 & 0.18 & 0.29 & 0.32\\
		 & \textsf{ArgumentPrediction} & \name{} & 0.64  & 0.86 & 0.92 & 0.52 & 0.66 & 0.70 & 0.57 & 0.75 & 0.80\\
		 &  & NL2Type & 0.53  & 0.80 & 0.88 & 0.46 & 0.61 & 0.66 & 0.50 & 0.70 & 0.76\\
		 &  & DeepTyper & 0.54  & 0.80 & 0.87 & 0.42 & 0.57 & 0.62 & 0.47 & 0.67 & 0.73\\
		 &  & Na\"ive baseline & 0.08  & 0.15 & 0.18 & 0.17 & 0.31 & 0.35 & 0.11 & 0.20 & 0.23\\
		\midrule
		\textsf{OSS} & \textsf{ReturnPrediction} & \name{} & 0.69  & 0.80 & 0.84 & 0.61 & 0.70 & 0.72 & 0.65 & 0.75 & 0.78\\
		 &  & NL2Type & 0.61  & 0.74 & 0.79 & 0.55 & 0.64 & 0.68 & 0.58 & 0.69 & 0.73\\
		 &  & DeepTyper & 0.52  & 0.79 & 0.83 & 0.48 & 0.59 & 0.64 & 0.50 & 0.66 & 0.72\\
		 &  & Na\"ive baseline & 0.16  & 0.25 & 0.28 & 0.25 & 0.42 & 0.47 & 0.20 & 0.31 & 0.35\\
		 & \textsf{ArgumentPrediction} & \name{} & 0.58  & 0.77 & 0.84 & 0.50 & 0.65 & 0.70 & 0.54 & 0.71 & 0.77\\
		 &  & NL2Type & 0.50  & 0.71 & 0.79 & 0.46 & 0.61 & 0.66 & 0.48 & 0.66 & 0.72\\
		 &  & DeepTyper & 0.51  & 0.76 & 0.84 & 0.45 & 0.59 & 0.64 & 0.48 & 0.67 & 0.73\\
		 &  & Na\"ive baseline & 0.06  & 0.11 & 0.14 & 0.14 & 0.25 & 0.29 & 0.08 & 0.15 & 0.19\\
		\bottomrule
	\end{tabular}
	\end{small}
  \vspace{-1em}
\end{table*}

\paragraph{Metrics}
\label{sec:metrics} We evaluate the effectiveness of \name{}'s neural type predictor by splitting the already annotated types in a given dataset into training (80\%) and validation (20\%) data.
The split is by file, to avoid mixing up types within a single file.
Once trained on the training data, we compare the model's predictions against the validation data, using the already annotated types as the ground truth.
We compute precision, recall, and F1-score, weighted by the number of type occurrences in the dataset.
Similarly to previous work~\cite{Malik2019}, if the prediction model cannot predict a type for a type slot (i.e., returns ``unknown''), we remove this type slot from the calculation of precision.
Specifically, we calculate precision as $prec = \frac{n_{corr}}{n_{all}}$, where $n_{corr}$ is
the number of correct predictions and $n_{all}$ is the number of type slots for which the
model does not return ``unknown''.
We calculate recall as $rec = \frac{n_{corr}}{|D|}$, where $|D|$ is total number of type slots in the examined dataset.
We report the \topk{} scores, for $k \in \{1,3,5\}$.

\paragraph{Baseline models}
We compare \name{}'s \topk{} predictions against three baseline models.
The \emph{na\"ive baseline} model considers the ten most frequent types in the dataset and samples its prediction from the distribution of these ten types, independently of the given context.
For example, it predicts \texttt{None} as a return type more often than \texttt{List[str]} because \texttt{None} is used more often as a return type than \texttt{List[str]}.
The \emph{DeepTyper} baseline is a Python re-implementation of the DeepTyper~\cite{Hellendoorn2018} model.
DeepTyper learns to translate a sequence of source code tokens to a sequence of types (and zeros for tokens without a type).
To make it directly compatible with \name{}, we do not consider predictions for variable annotations in function bodies, even though we do perform basic name-based type propagation in case an annotated argument is used in a function body.
Finally, the \emph{NL2Type} baseline is a re-implementation of the NL2Type model~\cite{Malik2019} for Python, which also learns from natural language information associated with a type, but does not consider code context or available types.
%We pick NL2Type and DeepTyper as the state-of-the-art baselines because they have been shown to outperform prior work on predicting types,
%including JSNice~\cite{Raychev2015}, albeit for a different language.

\paragraph{Results} \Cref{tab:effectiveness_model} presents the results for RQ~1.
Our neural model achieves moderate to high precision scores, e.g., 73\% in the
top-1 and 92\% in the top-5 for on the internal dataset for the
\textsf{ReturnPrediction} task.
The recall results are good but less high than precision, indicating that
\name{} is fairly confident when it makes a prediction, but abstains from doing so when it is not.
All models have slightly worse performance on the \textsf{OSS} dataset, which we attribute to the smaller size of that dataset.
The fact that the top-3 and top-5 scores are significantly higher than top-1 in all cases
motivates our work on combinatorial search (Section~\ref{sec:eval search}).

Compared to the baselines, \name{} outperforms both the state of the art and the na\"ive baseline across all metrics for both datasets and all three prediction tasks.
The differences between \name{} and NL2Type are higher in the case of the \textsf{ReturnPrediction} than the \textsf{ArgumentPrediction} task.
The context information, as obtained by analyzing token sequences, is helping the
\name{} prediction model more in the \textsf{ReturnPrediction} task.
Compared to DeepTyper, both NL2Type and \name{} are better, the latter by a
significant margin, in top-1 but not in top-3 or top-5 predictions.
Given that all models learn primarily from identifier names, the relatively
close upper bound performance seems to indicate that performance improvements
may only be achieved by introducing different (e.g., structural) information
to the model.

\subsection{RQ~2: Comparison with Simpler Variants of the Neural Model}

\begin{figure}
	\includegraphics[width=\linewidth]{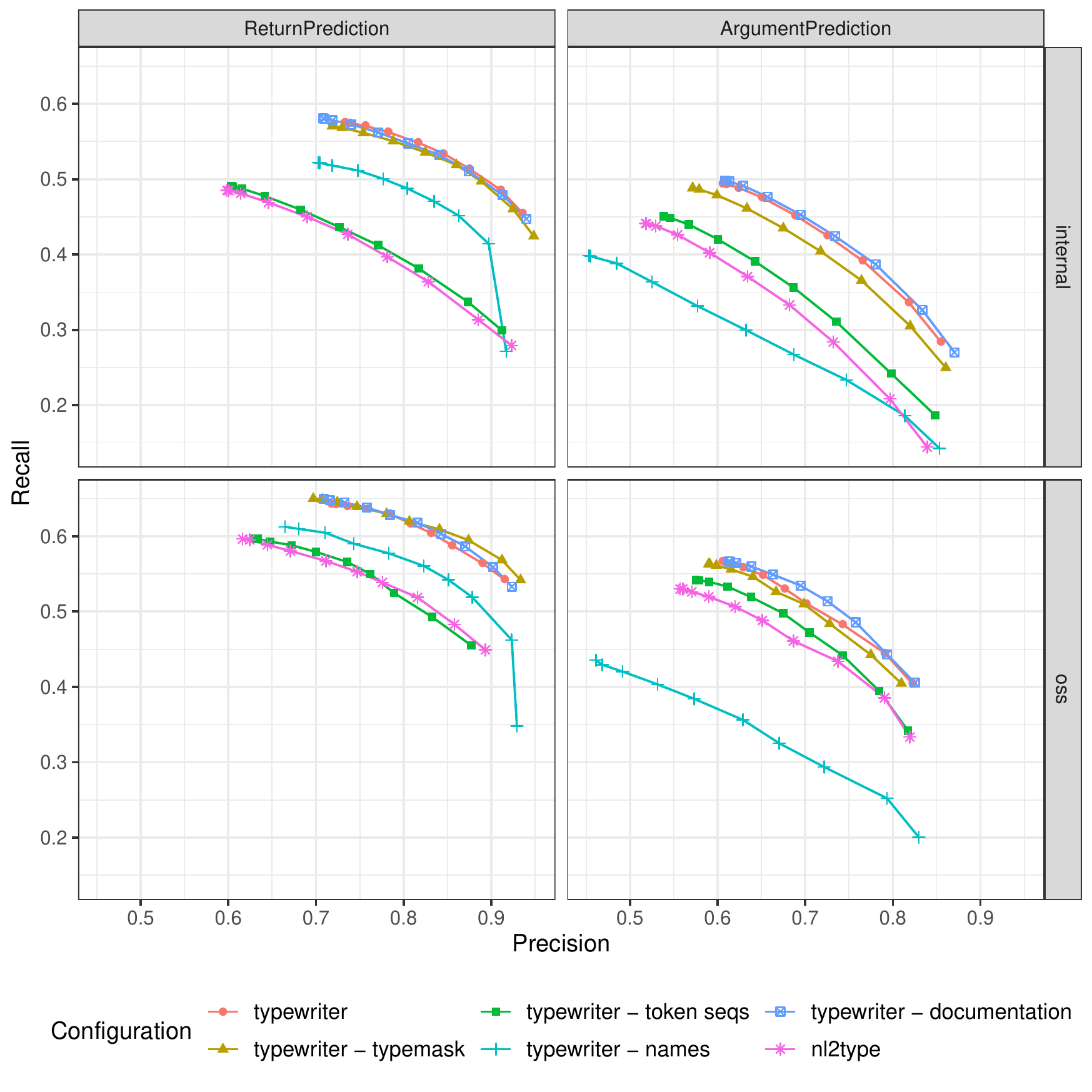}
	\caption{Precision/Recall curves for different \name configurations. Each data point represents a prediction threshold level.}
	\label{fig:ablation}
\end{figure}

The main novelty of \name{}'s prediction component is the inclusion of code context information and a local type mask in the prediction model.
%We expect code context to help the model learn from the usage locations of arguments in order to generalize beyond what function signatures and/or documentation can provide.
%We also expect that information about the types available in context will help the model learn to restrict predictions to types visible to the current prediction context.
To explore the influence of the different type hints considered by \name{}, we perform an ablation study. Specifically, we turn off parts of the model, both in training and in testing, and then measure top-1 precision and recall at various prediction threshold levels.
We start with the full model (\textsf{typewriter}) and then we remove, in order, the type mask, the token sequences, the method and argument names and the documentation. As a baseline, we also include \textsf{nl2type}, a configuration functionally equivalent with NL2Type~\cite{Malik2019},
which corresponds to \name{} without token sequences and without a type mask.
The results of the ablation study can be seen in~\Cref{fig:ablation}.

%\begin{itemize}
%
%	\item \textsf{typewriter}: This represents the proposed model in its full specification.
%
%	\item \textsf{typewriter - typemask}: \name{} without information about the available types
%	that are in context.
%
%	\item \textsf{typewriter - token seqs}: \name{} without information about the argument usage
%	locations.
%
%	\item \textsf{typewriter - names}: \name{} without information about the names of the
%	arguments and the function it tries to predict the types for.
%
%	\item \textsf{typewriter - docstrings}: \name{} excluding function documentation
%        (referred to as docstring in Python).
%
%	\item \textsf{nl2type}: This model is functionally equivalent with NL2Type~\cite{Malik2019}.
%	It corresponds to \name{} without token sequences and without a type mask. We use it here
%	as a baseline.
%
%\end{itemize}

Overall, the combined information of natural language, token sequences, and type masks helps \name{} to perform better than previous models, such as NL2Type.
The main contributor to this improvement is the token sequences component.
Moreover, the results seem to reinforce the main thesis of NL2Type, i.e., that natural language information and types are strongly related:
If we remove the argument and function naming information from \name{}, its performance drops significantly.

Contrary to our initial expectations, the type mask component is
not contributing significantly in the \textsf{ReturnPrediction} task, while only slightly improving
the \textsf{ArgumentPrediction} results.
We attribute this to the current implementation of the type mask data extraction process:
the extractor currently neither performs an in-depth dependency resolution to retrieve the full
set of types available in the processed file's name space, nor does it track type renamings (e.g., \texttt{import pandas as pd}).
The low predictive capability of comments can be explained by the fact that only
a small number of the methods in both datasets have documentation at the method level.
%It is interesting that for the \textsf{ArgumentPrediction} task, documentation strings actually
%slightly decrease the performance at high prediction confidence levels.
%At least in the case of Python, future versions of \name{} could be simplified by excluding
%documentation strings and type information from the prediction model.

\subsection{RQ~3: Effectiveness of Search}
\label{sec:eval search}

To evaluate the search, we collect a ground truth of 50 \emph{fully annotated} files that are randomly sampled from the industrial code base at \fb{}. We ensure that they type-check correctly.
The files we select originate from different products and vary in size and complexity, the files average 7 (median: 6, 95\%: 13) annotations.
The total number of annotations is 346.
%, of which 112 are out of vocabulary for our model (we exclude those from the results).
%
For each file in the ground truth, we strip its existing annotations and then apply \name{} to predict and evaluate the missing types.
We configure both the greedy and the non-greedy search strategies to stop when the number of states explored is seven times the number of type slots.
This threshold empirically strikes a reasonable balance between investing time and detecting correct types.
We use the same prediction model trained on the \fb{} dataset as in \Cref{sec:effectnn}.

\Cref{tab:effectiveness_search} shows the results on two levels: individual type annotations and files.
On the annotation-level, column \emph{type-correct} shows how many type slots the type assignment returned by the search fills (recall that the search ensures each added type to be type-correct).
Column \emph{ground truth match} shows how many of all added annotations match the original, developer-produced type annotations.
On the file-level, a \emph{complete and type-correct solution} is a file that \name{} fully annotates without type errors.
This metric does not include files where \name{} discovers a type-correct, but partially annotated solution.
The \emph{ground truth match} is the subset of the complete and type-correct solutions,
where the solution is identical to the ground truth for all types in the file.
%
% On the annotation-level, we compute precision and recall similarly to \Cref{sec:metrics}, by
% comparing the search results against the ground truth.
% On the file-level, we report the number of files for which the search strategy could find a solution that is type-correct, and the number of files for which the found solution is identical to the ground truth.
% We make this distinction as a type-correct solution may not necessarily match the ground truth.
% On the file-level, we report the number of files for which the search strategy could find a solution that is fully annotated and type-correct, and the number of files for which the found solution is identical to the ground truth.
% We make this distinction as a type-correct solution may not necessarily match the ground truth.
% On the annotation-level, we compute, from the files for which the search strategy finds a type-correct solution (but not necessarily fully annotated), the number of annotations
% discovered as well as the number of ground truth matches.
%
It is possible to find a type-correct annotation that does not match the ground truth.
For example, \name{} may correctly annotate the return type of a function as a \texttt{List}, but a human expert might choose a more precise type \texttt{List[str]}: both are type-correct, but the human annotation provides more guarantees.

% \begin{table}
%     \caption{Effectiveness of various search strategies for type inference.}
%     \label{tab:effectiveness_search}
%     \begin{small}
%     \resizebox{\columnwidth}{!}{
%         \begin{tabular}{p{0.14\columnwidth}cp{0.1\columnwidth}p{0.1\columnwidth}cp{0.14\columnwidth}p{0.14\columnwidth}cp{0.1\columnwidth}}
% 	\toprule
% 	Strategy & \topk & \multicolumn{2}{l}{Annotations} & \qquad & \multicolumn{2}{l}{Files} & \qquad & Time \\
%         \cmidrule{3-4} \cmidrule{6-7} \cmidrule{9-9}
% 	& & Precision & Recall & & type-correct solution & ground truth match & & mean sec per type \\
% 	\midrule
% 	\multirow{3}{0.1\columnwidth}{Prediction only} & 1 & 0.83 & 0.83 & & 28 (40\%) & 21 & & 0.2 \\
% 		                                       & 3 & 0.92 & 0.92 & & --- & 26 & & 0.2 \\
% 		                                       & 5 & 0.94 & 0.95 & & --- & 29 & & 0.2 \\
% 	\midrule
% 	\multirow{3}{0.1\columnwidth}{Greedy search} & 1 & 0.82 & 0.78 & & 28 (40\%) & 21 (30\%) & & 10\\
% 			                          & 3 & 0.86 & 0.86 & & 60 (85\%) & 23 (32\%) & & 12\\
% 		                                  & 5 & 0.87 & 0.86 & & 61 (87\%) & 23 (32\%) & & 12\\
% 	\midrule
% 	\multirow{3}{0.1\columnwidth}{Non-greedy search} & 1 & 0.83 & 0.79 & & 28 (40\%) & 21 (30\%) & & 50\\
% 			& 3 & 0.84 & 0.83 & & 68 (97\%) & 22 (31\%) & & 37\\
% 			& 5 & 0.84 & 0.83 & & 70 (100\%) & 22 (31\%) & & 37\\
% 	\midrule
% 	Pyre Infer & --- & 0.42 & 0.23 & & 12 & 10 &  & 1 \\
% 	\bottomrule
%         \end{tabular}
%     }
%     \end{small}
% \end{table}

\begin{table}
    \caption{Effectiveness of various search strategies for type inference.}
    \label{tab:effectiveness_search}
    \begin{small}
    \resizebox{\columnwidth}{!}{
        \begin{tabular}{p{0.14\columnwidth}cp{0.14\columnwidth}p{0.14\columnwidth}cp{0.14\columnwidth}p{0.14\columnwidth}c}
  \toprule
  Strategy & Top-$k$ & \multicolumn{2}{l}{Annotations} & \qquad & \multicolumn{2}{l}{Files} & \qquad  \\
        \cmidrule{3-4} \cmidrule{6-7}
  & & Type-correct  & Ground truth match & & Complete, type-correct & Ground truth match &  \\
  \midrule
  \multirow{3}{0.1\columnwidth}{Greedy search} & 1 & 176 (51\%) & 155 (45\%) & & 7 (14\%) & 5 (10\%) \\
                                               & 3 & 213 (62\%) & 169 (49\%) & & 14 (28\%) & 10 (20\%)\\
                                               & 5 & 248 (72\%) & 188 (54\%) & & 22 (44\%) & 11 (22\%) \\
  \midrule
  \multirow{3}{0.1\columnwidth}{Non-greedy search} & 1 & 175 (51\%) & 149 (44\%) & & 7 (14\%) & 5 (10\%) \\
                                                   & 3 & 150 (43\%) & 109 (32\%) & & 11 (22\%) & 7 (14\%) \\
                                                   & 5 & 152 (44\%) & 109 (32\%) & & 15 (30\%) & 5 (10\%) \\
  \midrule
\multirow{3}{0.1\columnwidth}{Upper bound (prediction)} & 1 & -- & 192 (55\%) & & -- & 5(10\%) \\
& 3 & -- & 234 (68\%) & & --- & 13 (26\%)  \\
& 5 & -- & 240 (69\%) & & --- & 14 (28\%) \\
  \midrule
  Pyre Infer & --- & 100 (29\%)  & 82 (24\%) & & 3 (2\%) & 2 (2\%) \\
  \bottomrule
        \end{tabular}
    }
    \end{small}
\end{table}

Both search strategies successfully annotate a significant fraction of all types.
On the annotation-level, they add between 40\% and 63\% of all types in a type-correct way, out of which 28\% to 47\% match the ground truth, depending on the search strategy.
On the file-level, \name{} completely annotates 14\% to 44\% of all files, and 10\% to 22\% of all files perfectly match the developer annotations.
Comparing the two search strategies, we find that, at the annotation-level, greedy search discovers more type-correct annotations with top-3 and top-5 predictions, while non-greedy search actually finds fewer annotations.
This is due to the exponential increase in search space, which makes it less likely that the non-greedy search finds a type-correct solution.
In contrast, the results suggest that the greedy search explores a more promising part of the search space.
At the file-level, both search approaches provide more annotations and fully annotate more files as the number of available predictions per slot increases.
In the greedy case, a search using the top-5 results still improves the outcome significantly; this suggests the search strategy can efficiently leverage the neural model's moderate improvement when $k$ increases beyond $3$.
%
%% MP: Removing this as it is confusing. Either we given more details, or we just omit it.
%The choice between greedy and non-greedy search also affect efficiency: On average, the non-greedy search takes three times longer to complete than the greedy-search.
%We do not report exact times, since the integration with the type-checker itself can be further optimized.

To better understand how effective the search is, we also show how many ground truth-matching types the top-$k$ predictions include (``upper bound (prediction)'').
Note that these numbers are a theoretical upper bound for the search, which cannot be achieved in practice because it would require exhaustively exploring all combinations of types predicted within the top-$k$.
Comparing the upper bound with the results of the search shows that the search gets relatively close to the maximum effectiveness it could achieve.
For example, a top-$5$ exploration with greedy search finds a complete and type-correct solution that matches the ground truth for 11 files, while the theoretical upper bound is 14 files.
We leave developing further search strategies, e.g., based on additional heuristics, for future work.

Overall, the results show that a greedy search among \topk{} types can uncover more types when given more predictions, while also maintaining type correctness.
$k = 5$ provides the best annotation performance.
While a non-greedy search should not immediately be disregarded, it should be considered in terms of how exhaustive the developer will allow the search to be.
\subsection{RQ~4: Comparing with Static Type Inference}

\begin{table}
	\caption{Comparison of \name{} and a traditional, static type inference (\tool{pyre infer}).}
	\label{fig:pyre_comparison}
	\begin{small}
	\resizebox{\columnwidth}{!}{

	\begin{tabular}{lrr}
		\toprule
		& \bfseries{Top-5 (greedy)} & \bfseries{Top-5 (non-greedy)} \\
		\midrule
		Total type slots                      & 346 & 346 \\
		\dots added by \name{} only           & 166 & 95 \\
		\dots added by \tool{pyre infer} only & 18 & 43\\
		\dots added by both tools             & 82 & 57 \\
		\hspace{1.5em} \dots same prediction  & 63 & 44 \\
		\dots neither could predict           & 80 & 151\\
		\bottomrule
	\end{tabular}
	}
	\end{small}
\end{table}

\begin{figure}
	\includegraphics[width=.9\linewidth]{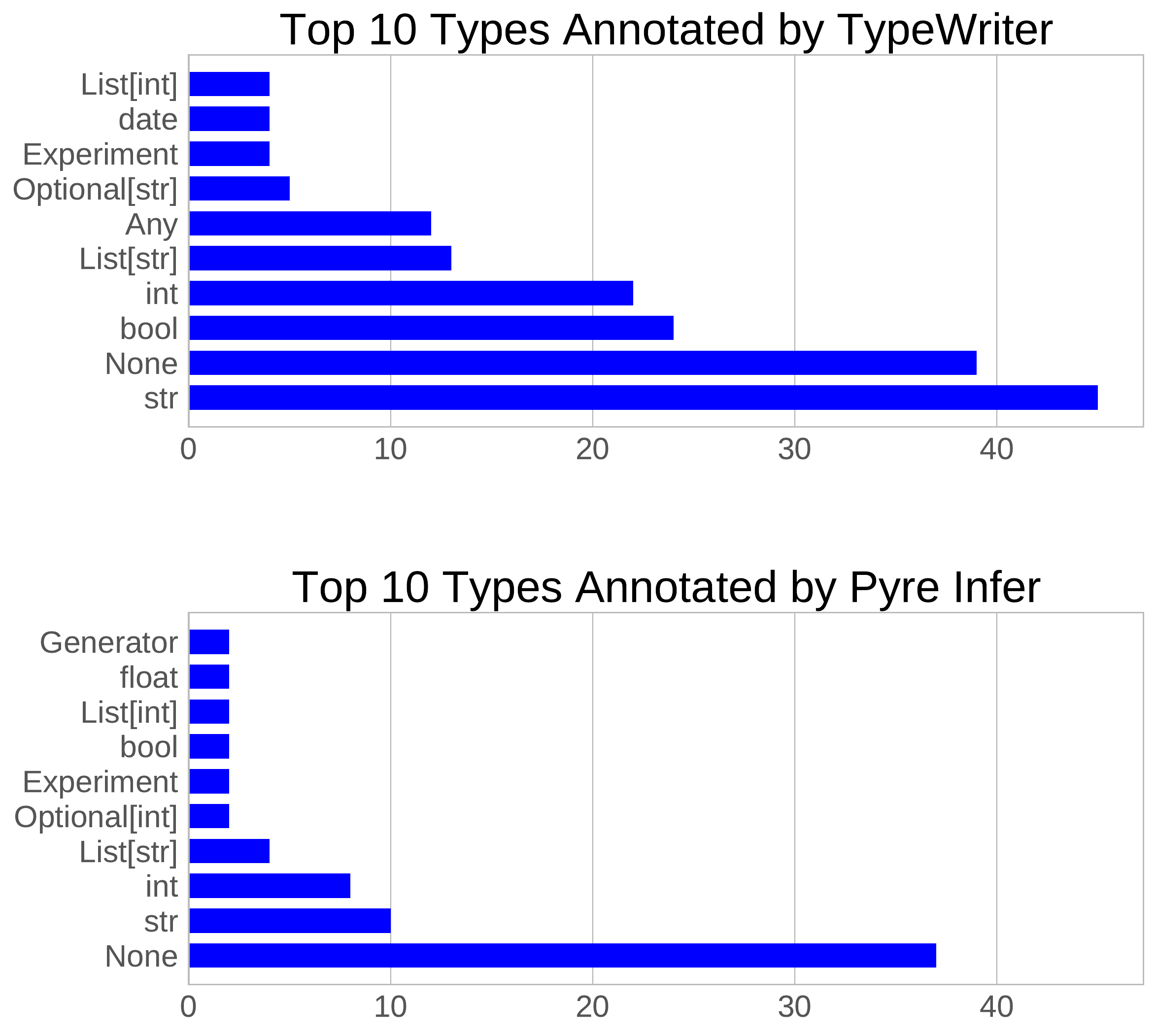}
	\caption{Distribution of types found by \name{} and Pyre Infer.}
	\label{fig:type_dist}
\end{figure}

We compare \name{} with a state-of-the-art, static type inference tool \tool{pyre infer}.
The type inference is part of the \tool{pyre} type checker and is representative of conservative static analysis-based type inference that adds only types guaranteed to be type-correct.
We run \tool{pyre infer} on the same set of randomly chosen, fully annotated files as in Section~\ref{sec:eval search} and then compare the added annotations with \name{}'s top-5 search results.

Tables~\ref{tab:effectiveness_search} (bottom) and~\ref{fig:pyre_comparison} show the results.
In a head to head comparison, \name{} is able to provide type-correct predictions for about seven times the number of files that \tool{pyre infer} can.
It also discovers significantly more types, adding a total of $188$ types, whereas \tool{pyre infer} adds only $100$.
Additionally, of the $82$ type slots for which both tools suggest a type, the suggestions are the same in $63$ cases.
Effectively, the types that \name{} suggests are a superset of those inferred by \tool{pyre infer}, as \tool{pyre infer} does not uniquely find many types.

To further illustrate the differences, we plot the distribution of the top-10 correctly predicted types in~\Cref{fig:type_dist}.
We see that \tool{pyre infer} can infer more precise types, but the majority of its inferences are on methods with no return types.
Moreover, some of the inferred types are of dubious usefulness (e.g., \texttt{Optional[Optional[Context]]}) indicating the difficulty of applying static type inference on dynamically-typed languages and reinforcing our thesis on the value of prediction-based type inference.

%\subsection{RQ~5: Efficiency of \name{}}
%
%A prediction model is only practical if it can be trained in reasonable type and its training
%can be scaled to large datasets. Moreover, when deployed, its runtime performance should be
%acceptable for interactive use cases.
%We evaluate \name{}'s efficiency by analyzing the time it takes to train it on our datasets
%and
%
%Training time
%
%Prediction time
%

\section{Discussion}

\paragraph{Effectiveness of neural type prediction} \name{} implements the first neural type prediction model for Python.
As all existing type prediction models~\cite{Malik2019, Hellendoorn2018, Raychev2015} target
JavaScript code, it is difficult to draw conclusions as to whether the \name{} architecture is the best for the task.
Two facts seem to suggest so: i) \name{} is better by a comfortable margin than
 a re-implementation of the two best-in-class JavaScript models, and ii) \name{}'s
performance is stable across two very different datasets.

\paragraph{Type-correctness vs.\ soundness}

Due to the way current Python type checkers work,
the types that \name{} produces are guaranteed to be \emph{type-correct} within the context of a given module.
Type correctness is different from type soundness, as the later can only be verified using human intuition.
This means that if a module is used within another context, the type checker might
invalidate an initially correct prediction.
In turn, this makes \name{} a \emph{soundy}~\cite{Livshits15}, rather than a sound approach.

\paragraph{Limited type vocabulary}

\name{} only predicts types that are part of its type vocabulary.
When the vocabulary size is configured at 1000 types, it can account for 90\% of the available types in both our datasets.
However, as software evolves, developers create new types or change the names of existing ones.
This may lead to situations where the model would predict a wrong type because its name changed
or because it simply does not know that the type exists.
The out-of-vocabulary problem is well know in software engineering
research~\cite{Hellen17}.
%In fact, Hellendoorn and Devanbu argue that it is the key challenge in applying deep models on code.
Recent work for by Karampatsis et al.~\cite{Karampatsis2020a} uses sub-word
information to account for neologisms with very good results.
We believe that \name{} would benefit significantly from such an approach for embedding identifier names, as it would enable it to learn semantically similar name variants (e.g., \texttt{AbstractClass} and \texttt{Class} or \texttt{List} and \texttt{List[str]}).

\paragraph{Further improvements} \name{} is a prototype stemming from a general
effort within \fb{} to make their Python code base more robust.
\name{} can be improved in several dimensions, some of which are presented
below:
%On an alpha testing basis, it has already been used to generate type annotations for individual files that have been accepted by developers.
%The following is a list of improvements that we are exploring before putting it into production use:

\subparagraph{\bfseries{Better data:}} The ablation study results suggest that
type masks and documentation components of the \name{} model are only
marginally contributing to its prediction capabilities. This goes against both intuition and published work: in~\cite{Malik2019}, the authors show that code
documentation is an important signal.
%Unfortunately, for both our datasets, the documentation coverage is not high.
We could, however, exploit the fact that highly used libraries, such as
\texttt{flask} or the Python standard library itself feature both type annotations (in
the \texttt{typeshed}\footnote{\href{https://github.com/python/typeshed/}{GitHub: python/typeshed}} repository) and excellent documentation.
Moreover, we can obtain better
type masks using lightweight dependency analysis, such as \tool{importlab},\footnote{\url{https://github.com/google/importlab}} to identify all types that are in context.
%Furthermore, teaching the model to learn how types are defined could allow it to automatically pick up all types associated with the existing context.

\subparagraph{\bfseries{Faster search feedback:}} \name{}'s execution speed is
currently constrained by the type checker used to obtain feedback. One natural way to
improve this would be to integrate the \name{} type predictor into a static
type inference loop: when the type inference cannot predict a type for a
location, it can ask the neural model for a suggestion. While the theoretical
cost of searching for types is similar, in practice the type inference
will be able to quickly examine suggested types given that all required data is loaded in memory.

\subparagraph{\bfseries{Reinforced learning:}} As with most neural models,
\name{} can benefit from more data. One idea worth exploring is to apply \name{} in batches,
consisting of application of an initial set of neural predictions, reviewing
proposed types through the normal code review process at \fb{} and then retrain
the model on the new data. At the scale of the \fb{} code base, we expect that
the feedback obtained (accepted, modified, and rejected suggestions) could be
used to improve the learning process.

\section{Related Work}

\paragraph{Type inference for dynamic languages}

Static type inference~\cite{Ander05, Furr09, Jensen09, Hassan18} computes types using, e.g., abstract interpretation or type constraint propagation.
These approaches are sound by design, but due to the dynamic nature of some languages, they
often infer only simple or very generic types~\cite{Furr09, Jensen09}.
They also require a significant amount of context, usually a full program and its dependencies.
Dynamic type inference~\cite{An11, Ren13} tracks data flows between functions, e.g., while executing a program's test suite.
These approaches capture precise types, but they are constrained by coverage.
\name{} differs from those approaches in two key aspects: i) it only requires limited context information, i.e., a single a source code file, ii) it does not require the program to be executed and hence can predict types in the absence of a test suite or other input data.

\paragraph{Probabilistic type inference}

The difficulty of accurately inferring types for dynamic programming languages has led to research on probabilistic type inference.
JSNice~\cite{Raychev2015} models source code as a dependency network of known (e.g., constants, API methods) and unknown facts (e.g., types); it then mines information from large code bases to quantify the probability of two items being linked together.
Xu et al.~\cite{Xu2016} predict variable types based on a probabilistic combination of multiple  uncertain type hints, e.g., data flows and attribute accesses.
They also consider natural language information, but based on lexical similarities of names, and focus on variable types, whereas \name{} focuses on function types.
DeepTyper~\cite{Hellendoorn2018} uses a sequence-to-sequence neural model to predict types based on a bi-lingual corpus of TypeScript and JavaScript code.
NL2Type~\cite{Malik2019} uses natural language information.
Our evaluation directly compares with Python re-implementations of both
DeepTyper and NL2Type.
Besides advances in the probabilistic type prediction model itself, the more important contribution of our work is to address the imprecision and combinatorial explosion problems of probabilistic type inference.
In principle, any of the above techniques can be combined with \name{}'s search-based validations to obtain type-correct types in reasonable time.

\paragraph{Type checking and inference for Python}

The Python community introduced a type annotation syntax
along with a type checker (\mypy{}) as part of Python 3.5 version in 2015~\cite{Rossum14}.
The combination of the two enables \emph{gradual typing} of existing code, where
the type checker checks only the annotated parts of the code.
Similar approaches have also been explored by the research community~\cite{Vitou15}.
%To cope with the fact that the type checker may not be able to infer the types of external libraries, the Python foundation maintains a GitHub repository (\texttt{python/typeshed}) with type-annotated function signatures for highly popular libraries, such as \texttt{scikit-learn} and \texttt{numpy}.
%
Since 2015, type annotations have seen adoption in several large-scale Python code bases, with products such as
Dropbox\footnote{\href{https://blogs.dropbox.com/tech/2018/09/how-we-rolled-out-one-of-the-largest-python-3-migrations-ever/}{Dropbox Blog: How we rolled out one of the largest Python 3 migrations ever}}
and Instagram,\footnote{\href{https://instagram-engineering.com/let-your-code-type-hint-itself-introducing-open-source-monkeytype-a855c7284881}{Instagram Engineering Blog: Introducing open source MonkeyType}} reportedly having annotated large parts of their multi-million line code bases.
\name{} helps reduce the manual effort required for such a step.

\paragraph{Machine learning on code}

Our neural type prediction model is motivated by a stream of work on machine learning-based program analyses~\cite{Allamanis2018}.
Beyond type prediction, others have proposed learning-based techniques to find programming errors~\cite{oopsla2018-DeepBugs,Nguyen2019}, predict variable and method names~\cite{Allamanis2015,Raychev2015,Vasilescu2017}, suggest how to improve names~\cite{Liu2019}, search code~\cite{Gu2018,Sachdev2018}, detect clones~\cite{White2016,Zhao2018}, classify code~\cite{Mou2016,Zhang2019}, predict code edits~\cite{Zhao2018b,Yin2018,Tufano2019}, predict assertions~\cite{Watson2020}, and automatically fix bugs~\cite{Gupta2017,Harer2018,oopsla2019}.
\name{} contributes a novel model for predicting types and a search-based combination of predictive models with traditional type checking.

\paragraph{Search-based software engineering}
Our search-based validation of types fits the search-based software engineering theme~\cite{harman2001search}, which proposes to balance competing constraints in developer tools through metaheuristic search techniques.
In our case, the search balances the need to validate an exponential number of combinations of type suggestions with the need to efficiently annotate types.

%Several type checkers, such as
%\tool{pyre},\footnote{GitHub: \href{https://github.com/facebook/pyre-check}{facebook/pyre-check}}
%\tool{pytype},\footnote{GitHub: \href{https://github.com/google/pytype}{google/pytype}} and
%\tool{pyright}\footnote{GitHub: \href{https://github.com/microsoft/pyright}{microsoft/pyright}} able to scale to very large code bases have emerged.

%Type checking can only work on type annotated code bases; to help developers annotate large
%code bases, tools such as
%\tool{monkeytype}\footnote{GitHub: \href{https://github.com/Instagram/MonkeyType}{Instagram/MonkeyType}},
%\tool{pysonar2}\footnote{GitHub: \href{https://github.com/yinwang0/pysonar2}{yinwang0/pysonar2}} and
%\tool{pyre infer} (part of \tool{pyre}) can automatically infer type annotations, using dynamic (\tool{monkeytype}) and static (\tool{pysonar2}, \tool{pyre infer}) techniques, respectively.

\section{Conclusions}

We present \name{}, a learning-based approach to the problem of inferring types for code written in Python. \name{} exploits the availability of partially annotated source code to learn a type
prediction model and the availability of type checkers to refine and validate the predicted types.
\name{}'s learned model can readily predict correct type annotations for half of the type slots on first try, whereas its search component can help prevent annotating code with wrong types.
Combined, the neural prediction and the search-based refinement helps annotate
large code bases with minimal human intervention, making \name{} the first practically applicable learning-based tool for type annotation.

We are currently in the process of making \name{} available to developers at \fb{}. We have tested the automation of the tool in the code review domain. Developers at \fb{} received type suggestions as comments on pull requests they had authored. They would also receive pull requests containing type suggestions for their project. The initial experience from applying the approach on a code base that powers tools used by billions of people has been positive: several thousands of suggested types have already been accepted with minimal changes.

\bibliographystyle{ACM-Reference-Format}
\bibliography{references}

\end{document}